\documentclass[twocolumn,showpacs,preprintnumbers]{revtex4}
\usepackage{amsmath}
\usepackage{graphicx}
\usepackage{dcolumn}
\usepackage{bm}

\input{tcilatex}

\begin{document}

\title{\bigskip Generation of a High-Visibility Four-Photon Entangled State
and Realization of a Four-Party Quantum Communication Complexity Scenario}
\author{Jin-Shi Xu, Chuan-Feng Li$\footnote{
email: cfli@ustc.edu.cn}$, and Guang-Can Guo}
\affiliation{Key Laboratory of Quantum Information, University of Science and Technology
of China, CAS, Hefei, 230026, People's Republic of China}
\date{\today }

\begin{abstract}
We obtain a four-photon polarization-entangled state with a visibility as
high as $(95.35\pm 0.45)\%$ directly from a single down-conversion source. A
success probability of $(81.54\pm 1.38)\%$ is observed by applying this
entangled state to realize a four-party quantum communication complexity
scenario (QCCS), which comfortably surpass the classical limit of $50\%$. As
a comparison, two Einstein-Podolsky-Rosen (EPR) pairs are shown to implement
the scenario with a success probability of $(73.89\pm 1.33)\%$. This
four-photon state can be used to fulfill decoherence-free quantum
information processing and other advanced quantum communication schemes.
\end{abstract}

\maketitle


Entanglement is one of the most important and interesting characteristic of
quantum mechanics. Entangled states of two or more particles not only play a
central role in the discussion of quantum mechanics versus local realism
\cite{1.Greenberger}, but also form the basis of nearly all quantum
information protocols, including quantum cryptography \cite{2.Tittel},
quantum computation \cite{3.Deutsch}, dense coding \cite{4.Mattle},
teleportation \cite{5.Bennett} and quantum communication complexity \cite%
{6.Cleve,7.Xue,8.Xue}.

Many experiments employing type-II spontaneous parametric
down-conversion (SPDC) process have been reported to realize
multiphoton entangled states, including a four-photon
Greenberger-Horne-Zeilinger (GHZ) state with a visibility of $(79\pm
6)\%$ \cite{9.Pan}, a four-photon decoherence-free state of
visibility $(79.3\pm 1.4)\%$ \cite{10.Eibl} and a four-photon
cluster state with a fidelity of $(74.1\pm 1.3)\%$ \cite{11.Kiesel}.
Yet, in those schemes interference occurs pairwise between processes
where the photon pair is created at distances $\pm x$ from the
middle of the crystal \cite{12.Kwiat}, which may limit the purity of
the state.

In this letter, we show that a polarization-entangled state observed behind
a single pulsed type-I SPDC source can reach a visibility as high as $%
(95.35\pm 0.45)\%$. We use this state to realize a four-party quantum
communication complexity scenario (QCCS) with a success probability of $%
(81.54\pm 1.38)\%$, which is much higher than the classical limit of $50\%$.
According to \v{C}. Brukner $et$ $al$. \cite{13.Brukner} this is equal to
show that our state violate a kind of Bell's inequality.

There is a reasonable probability of simultaneously producing four photons
in a single strong pulsed SPDC source. In our experiment, we use two
identically cut type-I beta-barium-borate (BBO) crystals (8.0$\times $8.0$%
\times $0.6 mm, $\theta _{pm}=30.35^{\text{o}}$) with their optic axes
aligned in mutually perpendicular planes \cite{14.Kwiat}. Frequency doubled
ultraviolet (UV) pluses (390 nm center wavelength, $\sim $200 fs pulse
duration, 76 MHz repetition rate, $\sim $500 mW average power) from a
mode-locked Ti:sapphire laser is used to pass through the two-crystal
geometry BBO. Behind two 50-50 beam splitters, the four photons with
distinct spatial mode are coupled into single mode optical fibers (Fig. 1).
\begin{figure}[tbph]
\begin{center}
\includegraphics[width= 3.3in]{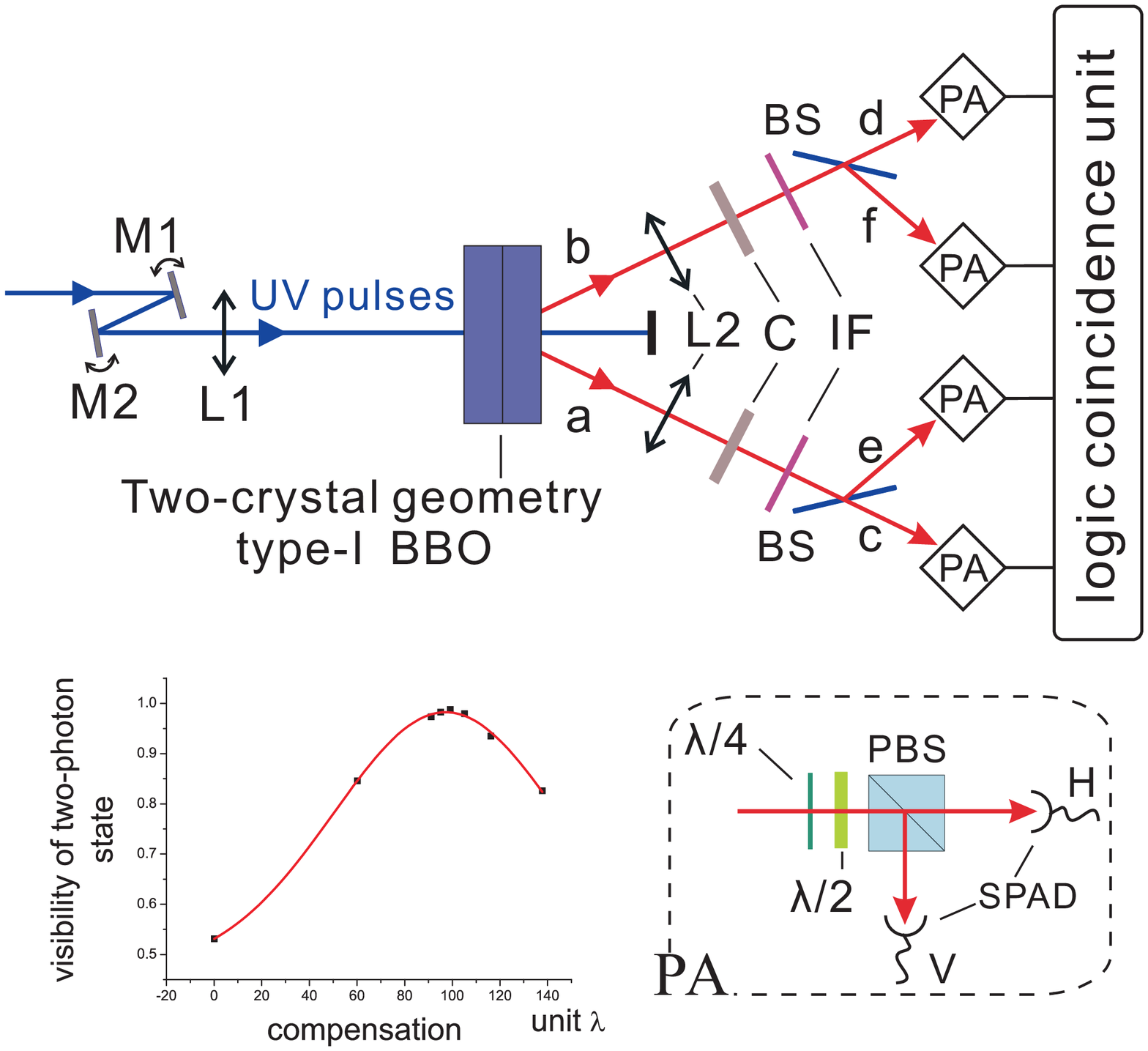}
\end{center}
\caption{(Color on line). Experimental setup. The UV pulse is focused by a
convex lens with a focal length of 50 cm (L1) and the direction of the UV
beam is controlled by two mirrors M1 and M2. The four photons are emitted
into two spatial modes $a$ and $b$. After collected by a convex lens with a
focal length of 30 cm (L2) in each mode, the four photons pass through
quartz plates (C) to compensate the birefringence in BBO. Then, they are
distributed into the four modes $c$, $d$, $e$ and $f$ by two 50-50 beam
splitters (BS) behind interference filters (IF, $\Delta \protect\lambda =3$
nm, $\protect\lambda =780$ nm). In order to analyze the four-photon state
and to realize the QCCS, polarization analysis (PA) in various bases is
performed for each mode using quarter wave plates $(\protect\lambda /4)$\
and half wave plates $(\protect\lambda /2)$\ in front of polarizing beam
splitters (PBS) and single photon avalanche detectors (SPAD). The inset
shows the visibility of two-photon entangled state versus compensation (100
mW pump). The solid line is a Gaussian function fitting (unit $\protect%
\lambda =780$ nm).}
\end{figure}

According to Schr\"{o}dinger equation, the four-photon state can be obtained
as
\begin{eqnarray}
\left\vert \Psi ^{4}\right\rangle &=&(a_{H}^{\dag }b_{H}^{\dag }+a_{V}^{\dag
}b_{V}^{\dag })^{2}\left\vert 0\right\rangle  \notag \\
&=&(a_{H}^{\dag 2}b_{H}^{\dag 2}+a_{V}^{\dag 2}b_{V}^{\dag 2}+2a_{H}^{\dag
}b_{H}^{\dag }a_{V}^{\dag }b_{V}^{\dag })\left\vert 0\right\rangle \text{,}
\label{1}
\end{eqnarray}%
where $a_{H}^{\dag }$ is the creation operator of a photon with horizontal
polarization in mode $a$, etc.

For simplicity, we assume that at the beam splitters $a$ is transformed into
$\frac{1}{\sqrt{2}}(c+e)$ and $b$\ into $\frac{1}{\sqrt{2}}(d+f)$ \cite%
{15.Weinfurter}, where $c$, $d$ and $e$, $f$ denote the transmitted and
reflected modes, respectively. We then expand Eq. (1) and keep only those
terms which lead to four-photon coincidence behind the two beam splitters,
i.e., only those terms for which there is one photon in each of the modes.
As a result, this four-photon state can be written as
\begin{eqnarray}
\ \left\vert \Psi ^{4}\right\rangle &=&\left\vert HHHH\right\rangle
_{cdef}+\left\vert VVVV\right\rangle _{cdef}  \notag \\
&&+\frac{1}{2}(\left\vert HHVV\right\rangle _{cdef}+\left\vert
HVVH\right\rangle _{cdef}  \notag \\
&&+\left\vert VHHV\right\rangle _{cdef}+\left\vert VVHH\right\rangle _{cdef})%
\text{,}  \label{4}
\end{eqnarray}%
where $\left\vert HHHH\right\rangle _{cdef}$ denotes a $H$ polarized photon
in each mode of $c$, $d$, $e$ and $f$, etc.

This state can be seen as the superposition of a four-photon GHZ state and a
product of two Einstein-Podolsky-Rosen (EPR) pairs (normalized)
\begin{equation}
\left\vert \Psi ^{4}\right\rangle =\sqrt{\frac{2}{3}}\left\vert
GHZ\right\rangle _{cedf}+\sqrt{\frac{1}{3}}\left\vert EPR\right\rangle
_{ce}\otimes \left\vert EPR\right\rangle _{df}\text{,}  \label{5}
\end{equation}%
where $\left\vert GHZ\right\rangle =\frac{1}{\sqrt{2}}(\left\vert
HHHH\right\rangle +\left\vert VVVV\right\rangle )$ is the GHZ state, $%
\left\vert EPR\right\rangle =\frac{1}{\sqrt{2}}(\left\vert HV\right\rangle
+\left\vert VH\right\rangle )$ is the EPR state $\left\vert \Psi
^{+}\right\rangle $.

Many efforts have been made to keep our experimental system stable for
several days. An air conditioner is used to keep the room temperature to the
order of $\pm 1^{\circ }$C. To avoid damage to the second harmonic
generation BBO and the SPDC BBO, we pump N$_{2}$\ around them. Moreover, by
using a motion controller system (Newport, NSC200) to tilt two mirrors M1
and M2 (in Fig. 1) with the feedback of two charge coupled devices (not
shown) and twofold coincidences of two paths (such as modes $c$ and $d$), we
manage to maintain the position of the pump beam.

To obtain the high-purity four-photon entangled state, the birefringence
between horizontal and vertical photons in the two-crystal geometry BBO has
been compensated with quartz plates. The inset of Fig. 1 shows that the
coherence between horizontal and vertical photons is recovered perfectly
while the compensation of optical path difference is about $99.1$ $\lambda $.

Fig. 2(a) and (b) show the 16 possible fourfold coincidence probabilities
for detecting one photon in each mode with the four polarization analyzers
oriented along $H/V$ basis\ and $+/-$ basis ($\pm 45^{\circ }$ linear
polarizations, i.e., $\frac{1}{\sqrt{2}}(H\pm V)$), respectively \cite%
{10.Eibl}. The integration time is 3 hours per column. One can find two
types of coincidences, the GHZ part, and the fourfold coincidences due to
the EPR pairs with average rates lower by a factor of 4, which is in very
good agreement with the state in Eq. (3).
\begin{figure}[tbph]
\begin{center}
\includegraphics[width= 3.3in]{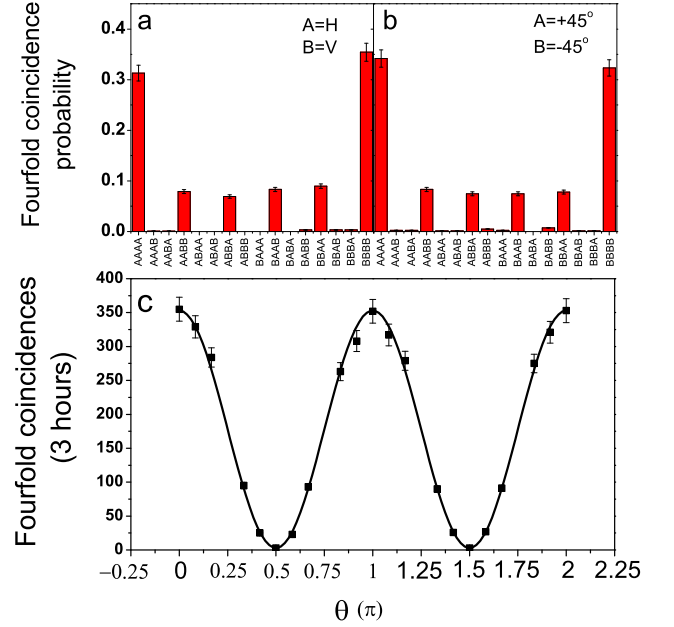}
\end{center}
\caption{(Color on line). Fourfold coincidence probabilities corresponding
to different measurement basis settings. (a) the $H/V$ basis; (b) the $+/-$
basis. (c) four-photon interference curve. We vary the detection basis in
mode $f$, while keeping mode $c$, $d$, $e$ in the $+$ ($+45^{\circ }$)
basis. $\protect\theta $ represent the angle between the linear polarization
detection basis and the $+$ basis in mode $f$. The solid line shows a
sinusoidal fit to the experimental results with a visibility of $(98.45\pm
0.15)\%$.}
\end{figure}

Fig. 2(c) shows one of the four-photon interference curves of the entangled
state. More strictly, we may use the correlation function to characterize
the entangled state. The experimental value of the correlation function is
obtained from the 16 four-photon coincidence rates with \cite{10.Eibl}
\begin{eqnarray}
\ E(\phi _{c},\phi _{e},\phi _{d},\phi _{f}) &=&\underset{%
l_{c},l_{e},l_{d},l_{f}=\pm 1}{\sum }l_{c},l_{e},l_{d},l_{f}  \notag \\
&&\times P_{l_{c},l_{e},l_{d},l_{f}}(\phi _{c},\phi _{e},\phi _{d},\phi _{f})%
\text{,}  \label{6}
\end{eqnarray}%
where $l_{x}$, $\phi _{x}$ are corresponding to the eigenvectors $\left\vert
l_{x},\phi _{x}\right\rangle =\frac{1}{\sqrt{2}}(\left\vert V\right\rangle
_{x}+l_{x}e^{-i\phi _{x}}\left\vert H\right\rangle _{x})$ with eigenvalues $%
l_{x}=\pm 1$ for polarization measurements performed by the observation
stations in the four modes $(x=c,e,d,f)$ and $P_{l_{c},l_{e},l_{d},l_{f}}$
are the four-photon\ probabilities. Theoretically \cite{15.Weinfurter}, when
$\phi _{c}=\phi _{e}=\phi _{d}=\phi _{f}=0$, the correlation function
reaches its maximal value, which is equal to the visibility of the curve of $%
E$ versus one of the angles, such as $\phi _{c}$, with other angles $\phi
_{e}=\phi _{d}=\phi _{f}=0$. From the data of Fig. 2(b), we obtain $%
V=(95.35\pm 0.45)\%$, compared to the theoretical result $V=100\%$ for a
pure state.

During the experiment, different setup is attempted to achieve better
result. For example, if the lens with a focal length of 50 cm is replaced by
a lens of 30 cm to focus the pump pulse onto the crystal, the four-photon
coincidences will be $4$ times brighter, however the four-photon visibility
will decrease to about $80\%$.

Next, we use this entangled state to realize a four-party QCCS.

Suppose there are four parties A, B, C and D receiving $X$, $Y$, $Z$ and $K$%
, respectively, where $X$, $Y$, $Z$, $K\in U\in \{0$, $1\}^{2}$, and they
are promised that\ \ \ \ \ \ \ \ \ \ \ \ \ \ \ \ \ \ \ \ \ \ \ \ \ \ \ \
\begin{equation}
(X+Y+Z+K)\func{mod}2=0\text{.}  \label{7}
\end{equation}%
The common goal is for each party to get the correct value of the Boolean
function\ \ \ \ \ \ \ \ \ \ \ \ \ \ \ \ \ \ \ \ \ \ \ \ \ \
\begin{equation}
\ F(X,Y,Z,K)=\frac{1}{2}[(X+Y+Z+K)\func{mod}4]\text{.}\   \label{8}
\end{equation}%
$X$, $Y$, $Z$ and $K$ can be represented in binary notation as $x_{1}x_{0}$,
$y_{1}y_{0}$, $z_{1}z_{0}$ and $k_{1}k_{0}$.\ According to Eq. (5)\ $%
x_{0}y_{0}z_{0}k_{0}$ is one of the eight combinations\ $0000$, $0011$, $%
0101 $, $0110$, $1001$, $1010$, $1100$, $1111$.$\ $\ \ \ \ \ \ \ \ \ \ \ \ \
\ \ \ \

We then rewrite Eq. (6) as\ \ \ \ \ \ \ \ \ \ \ \ \ \ \ \ \ \ \ \ \
\begin{equation}
\ \ F=x_{1}\oplus y_{1}\oplus z_{1}\oplus k_{1}\oplus
F_{0}(x_{0},y_{0},z_{0},k_{0})\text{,}  \label{9}
\end{equation}%
where%
\begin{equation}
F_{0}(x_{0},y_{0},z_{0},k_{0})=\frac{1}{2}[(x_{0}+y_{0}+z_{0}+k_{0})\func{mod%
}4]\text{.}  \label{10}
\end{equation}%
As a result, if $x_{0}y_{0}z_{0}k_{0}=0000$ $or$ $1111$, $F_{0}=0$, else, $%
F_{0}=1$.

Obviously, if these four parties are restricted to broadcast one bit
respectively, they have $50\%$ probability to get the correct value of $F$
in classical situation \cite{8.Xue}. On the other hand, if they share the
four-photon entangled state we have prepared initially, the probability they
get the correct value of $F$ can reach $83.33\%$, as shown below.

Each of the four parties A, B, C and D share one photon of the state\ \ \ \
\begin{eqnarray}
\left\vert \Psi ^{4}\right\rangle &=&\frac{1}{\sqrt{3}}(\left\vert
0000\right\rangle +\left\vert 1111\right\rangle )  \notag \\
&&+\frac{1}{2\sqrt{3}}(\left\vert 0011\right\rangle +\left\vert
1001\right\rangle +\left\vert 0110\right\rangle +\left\vert
1100\right\rangle )\text{, }  \notag \\
&&  \label{11}
\end{eqnarray}%
where $0$ and $1$\ represents $H$ and $V$ in Eq. (2), respectively. (we have
omitted the subscripts $c,d,e,f$ for simplicity).

If $x_{0}$ ($y_{0}$, $z_{0}$, $k_{0}$)$=0$, then A (B, C, D) applies rotation

$\ \ \ \ \ \ \ \ \ \ \ \ \ \ \ \ \ \ \ \ R(x)=\frac{1}{\sqrt{2}}\left(
\begin{array}{cc}
1 & 1 \\
1 & -1%
\end{array}%
\right) $

\noindent on his own photon with half wave plate and quarter wave plate; if $%
x_{0}$ ($y_{0}$, $z_{0}$, $k_{0}$)$=1$, then A (B, C, D) applies rotation

$\ \ \ \ \ \ \ \ \ \ \ \ \ \ \ \ \ \ \ \ R(y)=\frac{1}{\sqrt{2}}\left(
\begin{array}{cc}
1 & i \\
i & 1%
\end{array}%
\right) $

\noindent on his own photon. Then each of the four parties measures the
photon under $0/1$ $(H/V)$ basis and get the result of $a,b,c,d$. Due to the
entanglement of the state they share initially, A, B, C, D only have to
broadcast the four bits $x_{1}\oplus a$, $y_{1}\oplus b$, $z_{1}\oplus c$, $%
k_{1}\oplus d$, respectively. Then they have on average $83.33\%$
probability to get the correct value of $F$ as
\begin{equation}
\ F=x_{1}\oplus a\oplus y_{1}\oplus b\oplus z_{1}\oplus c\oplus k_{1}\oplus d%
\text{,}  \label{12}
\end{equation}%
that is \ \ \ \ \ \ \ \ \ \ \ \ \ \ \ \ \ \ \ \ \ \ \ \ \ \ \ \ \ \
\begin{equation}
F_{0}=a\oplus b\oplus c\oplus d.  \label{13}
\end{equation}

For example, in the case $x_{0}y_{0}z_{0}k_{0}=0000$, local rotations$\
R(x)\otimes R(x)\otimes R(x)\otimes R(x)$\ do not change the four-photon
state, i.e., $\left\vert \Psi ^{4}\right\rangle ^{\prime }=\left\vert \Psi
^{4}\right\rangle $, where $\left\vert \Psi ^{4}\right\rangle ^{\prime }$ is
the state after local rotations. Consequently, the success probability is $%
100\%$. The remaining cases can be similarly analyzed, as shown in table I.

\begingroup\squeezetable
\begin{table*}[tbp]
\caption{The input of $x_{0}y_{0}z_{0}k_{0}$, the corresponding local
rotations, the components of \TEXTsymbol{\vert}$\Psi ^{4}\rangle ^{\prime }$
for successful communication, the result of $F_{0}$, the corresponding
theoretical probability (theor. prob.) and experimental probability (expt.
prob.) to get the correct value of $F$.}%
\begin{tabular}{|l|l|llll|l|l|l|}
\hline
$x_{0}y_{0}z_{0}k_{0}$ \ \ \  & \ \ \ \ \ \ \ Local rotations \ \ \ \ \ \ \
\ \ \  & \  & $\ $\TEXTsymbol{\vert}$\Psi ^{4}\rangle ^{\prime }$ &  &  & \ $%
\ \ F_{0}$ \ \ \  & \ \ \ Theor. prob. \ \ \  & \ \ \ \ \ \ \ Expt. prob. \
\ \ \ \ \  \\ \hline
$0000$ & $R(x)\otimes R(x)\otimes R(x)\otimes R(x)$ & $\left\vert
0000\right\rangle $ & $\left\vert 0011\right\rangle $ & $\left\vert
0101\right\rangle $ & $\left\vert 0110\right\rangle $ & \ $\ \ 0$ & $\ \ \ \
\ 100\%$ \ \  & $\ \ \ \ 97.68\%\pm 0.23\%$ \ \  \\ \cline{1-2}\cline{7-9}
$1111$ & $R(y)\otimes R(y)\otimes R(y)\otimes R(y)$ & $\left\vert
1001\right\rangle $ & $\left\vert 1010\right\rangle $ & $\left\vert
1100\right\rangle $ & $\left\vert 1111\right\rangle $ & $\ \ \ 0$ & $\ \ \ \
\ 100\%$ \ \  & $\ \ \ \ 96.32\%\pm 0.35\%$ \\ \hline
$0011$ & $R(x)\otimes R(x)\otimes R(y)\otimes R(y)$ &  &  &  &  & $\ \ \ 1$
& $\ \ \ \ 83.33\%$ \ \  & $\ \ \ \ 80.49\%\pm 1.57\%$ \\
\cline{1-2}\cline{7-9}
$1100$ & $R(y)\otimes R(y)\otimes R(x)\otimes R(x)$ & $\left\vert
0001\right\rangle $ & $\left\vert 0010\right\rangle $ & $\left\vert
0100\right\rangle $ & $\left\vert 0111\right\rangle $ & $\ \ \ 1$ & $\ \ \ \
83.33\%$ \  & $\ \ \ \ 82.63\%\pm 1.44\%$ \\ \cline{1-2}\cline{7-9}
$0110$ & $R(x)\otimes R(y)\otimes R(y)\otimes R(x)$ & $\left\vert
1000\right\rangle $ & $\left\vert 1011\right\rangle $ & $\left\vert
1101\right\rangle $ & $\left\vert 1110\right\rangle $ & $\ \ \ 1$ & $\ \ \ \
83.33\%$ & $\ \ \ \ 79.06\%\pm 1.66\%$ \\ \cline{1-2}\cline{7-9}
$1001$ & $R(y)\otimes R(x)\otimes R(x)\otimes R(y)$ &  &  &  &  & $\ \ \ 1$
& $\ \ \ \ 83.33\%$ & $\ \ \ \ 84.13\%\pm 1.34\%$ \\ \hline
$0101$ & $R(x)\otimes R(y)\otimes R(x)\otimes R(y)$ & $\left\vert
0001\right\rangle $ & $\left\vert 0010\right\rangle $ & $\left\vert
0100\right\rangle $ & $\left\vert 0111\right\rangle $ & $\ \ \ 1$ & $\ \ \ \
66.67\%$ & $\ \ \ \ 67.18\%\pm 2.20\%$ \\ \cline{1-2}\cline{7-9}
$1010$ & $R(y)\otimes R(x)\otimes R(y)\otimes R(x)$ & $\left\vert
1000\right\rangle $ & $\left\vert 1011\right\rangle $ & $\left\vert
1101\right\rangle $ & $\left\vert 1110\right\rangle $ & $\ \ \ 1$ & $\ \ \ \
66.67\%$ & $\ \ \ \ 64.82\%\pm 2.28\%$ \\ \hline
\end{tabular}%
\end{table*}
\endgroup
\begin{figure}[tbph]
\begin{center}
\includegraphics[width= 3.3in]{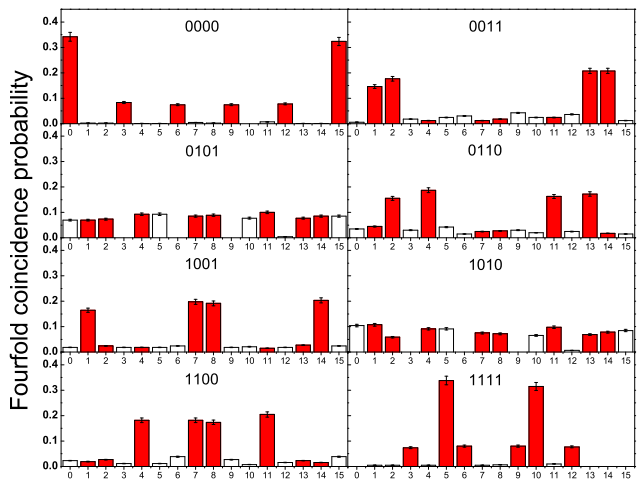}
\end{center}
\caption{(Color on line). Fourfold coincidence probabilities for the
four-party QCCS. Each frame represents a kind of combination of $%
x_{0}y_{0}z_{0}k_{0}$, denoted by $0000$, $0011$, etc. The $x$ axis of each
frame ($0$,$1$ $\cdot \cdot \cdot $ $15$) represents the sixteen different $%
0/1$ ($H/V$) basis settings in binary representation, e.g., 9=1001. The
filled and unfilled\ columns denote the probabilities of getting the correct
and wrong value of $F$, respectively.}
\end{figure}

Fig. 3 illustrates the experimental result in detail. It is shown that the
average probability for the four parities to get the correct value of $F$ in
our experiment is $(81.54\pm 1.38)\%$, which greatly surpass the classical
limit of $50\%$. This result prove that the state we have prepared violate a
kind of Bell's inequality \cite{13.Brukner}.

To illustrate that there is genuine four-photon entanglement in the state we
have prepared, we further consider another case, where (A, B) and (C, D)
share two EPR states $\left\vert \Phi ^{+}\right\rangle =\frac{1}{\sqrt{2}}%
(\left\vert HH\right\rangle +\left\vert VV\right\rangle )=\frac{1}{\sqrt{2}}%
(\left\vert 00\right\rangle +\left\vert 11\right\rangle )$, respectively. It
can be deduced that the probability for the parties to get the correct value
of $F$ is $75\%$ in this case.
\begin{figure}[tbph]
\begin{center}
\includegraphics[width= 3.3in]{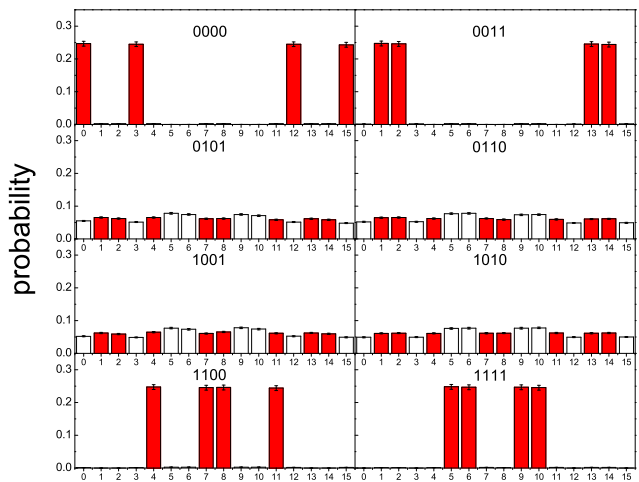}
\end{center}
\caption{(Color on line). Probability distribution for the four-party QCCS
when the four parties share two identical EPR states.}
\end{figure}

In experiment, The EPR state is generated with 100 mW UV pulses. We
manipulate (A,B) and (C,D) independently with twofold coincidences and
combine their results to get the probability distribution for the four-party
QCCS.

Fig. 4 shows the experimental result when these four parties share two EPR
states. The average success probability we obtain is $(73.89\pm 1.33)\%$.
Compared to the experimental result of the entangled state we have prepared,
we can see that it is the genuine four-photon entangled part ($\left\vert
GHZ\right\rangle $ in Eq. (3)) making the success probability reach $%
(81.54\pm 1.38)\%$.

In conclusion, we have prepared a four-photon entangled state with a
visibility as high as $(95.35\pm 0.45)\%$. By using this state to realize a
four-party QCCS, we prove that the state we have\ prepared violates a kind
of Bell's inequality indirectly. This four-photon state can be used to
fulfill decoherence-free quantum information processing \cite{16.Bourennane}
and other\ advanced quantum communication schemes.

This work was funded by National Natural Science Foundation of China.


\begin{thebibliography}{99}
\bibitem{1.Greenberger} D. M. Greenberger, M. A. Horne, A. Shimony, and A.
Zeilinger, Am. J. Phys. 58, 1131 (1990).

\bibitem{2.Tittel} W. Tittel, J. Brendel, H. Zbinden, and N. Gisin, Phys.
Rev. Lett. 84, 4737 (2000); T. Jennewein, C. Simon, G. Weihs, H. Weinfurter,
and A. Zeilinger, Phys. Rev. Lett. 84, 4729 (2000); D. S. Naik, C. G.
Peterson, A. G. White, A. J. Berglund, and P. G. Kwiat, Phy. Rev. Lett. 84,
4733 (2000).

\bibitem{3.Deutsch} D. Deutsch and R. Jozsa, Proc. R. Soc. London, Ser. A
439, 553 (1992).

\bibitem{4.Mattle} K. Mattle, H. Weinfurter, P. G. Kwiat, and A. Zeilinger,
Phys. Rev. Lett. 76, 4656 (1996).

\bibitem{5.Bennett} C. H. Bennett, G. Brassard, C. Cr\'{e}peau, R. Jozsa, A.
Peres, and W. K. Wootters, Phys. Rev. Lett. 70, 1895 (1993); D.
Bouwmeester, J.-W. Pan, K. Mattle, M. Eibl, H. Weinfurter, and A.
Zeilinger, Nature (London) 390, 575 (1997).

\bibitem{6.Cleve} R. Cleve and H. Buhrman, Phy. Rev. A 56, 1201 (1997).

\bibitem{7.Xue} P. Xue, Y.-F Huang, Y.-S Zhang, C.-F Li, and G.-C Guo, Phy.
Rev. A 64, 032304 (2001).

\bibitem{8.Xue} P. Xue, C.-F. Li, Y.-S Zhang, and G.-C Guo, J. Opt. B:
Quantum Semiclassical Opt. 3, 219 (2001).

\bibitem{9.Pan} J.-W. Pan, M. Daniell, S. Gasparoni, G. Weihs, and A.
Zeilinger, Phy. Rev. Lett. 86, 4435 (2001).

\bibitem{10.Eibl} M. Eibl, S. Gaertner, M. Bourennane, C. Kurtsiefer, M.
\.{Z}ukowski, and H. Weinfurter, Phy. Rev. Lett. 90, 200403 (2003).

\bibitem{11.Kiesel} N. Kiesel, C. Schmid, U. Weber, G. T\'{o}th, O. G\"{u}%
hne, R. Ursin, and H. Weinfurter, Phy. Rev. Lett. 95, 210502 (2005).

\bibitem{12.Kwiat} P. G. Kwiat, K. Mattle, H. Weinfurter, A.
Zeilinger, A. V. Sergienko, and Y. Shih, Phys. Rev. Lett. 75, 4337
(1995).

\bibitem{13.Brukner} \v{C}. Brukner, M. \.{Z}ukowski, J.-W Pan, and A.
Zeilinger, Phys. Rev. Lett. 92, 127901 (2004).

\bibitem{14.Kwiat} P. G. Kwiat, E. Waks, A. G. White, I. Appelbaum, and P.
H. Eberhard, Phys. Rev. A 60, R773 (1999).

\bibitem{15.Weinfurter} H. Weinfurter and M. \.{Z}ukowski, Phys. Rev. A 64,
010102(R) (2001).

\bibitem{16.Bourennane} M. Bourennane, M. Eibl, S. Gaertner, C. Kurtsiefer,
A. Cabello, and H. Weinfurter, Phys. Rev. Lett. 92, 107901 (2004).
\end{thebibliography}
\end{document}